# Lightning impulse current tests on some electroconductive fabrics

J. A. Cristancho[a*] • J. E. Rodriguez[a] • C. A. Rivera[a] • J. J. Pantoja[a,c] • L. K. Herrera[b] • F. Roman[a]

[a]*Electromagnetic Compatibility Research Group – EMC-UN, Facultad de Ingeniería, Universidad Nacional de Colombia*
[b]*Grupo de análisis de falla, integridad y superficie AFIS, Facultad de Ingeniería, Universidad Nacional de Colombia*
[c]*Technology Innovation Institute, Directed Energy Research Centre, Abu Dhabi, UAE*

**Abstract:** On the search of lightweight lightning protection materials that can be used as part of lightning protection systems (LPS), we investigate some types of electroconductive fabrics by applying several lightning impulse currents in laboratory. Samples of four commercially available electroconductive textiles were analyzed: two rip-stop, a plain-weave, a nonwoven, and additionally a carbon-impregnated polymeric film. Under laboratory conditions, each sample was subjected to several lengthwise subsequent lightning-like currents of 8/20 µs waveform, recording both voltage and current signals. Optical and scanning electron microscope observations were performed after tests, revealing some patterns or morphological changes on the fabric surface. Despite these changes, the investigated conductive textiles withstand the several lightning impulse currents applied. Results suggest that some conductive fabrics could be used in personal mobile shelters, to protect human beings against the earth's potential rise caused by a close lightning discharge.

*Corresponding author.
*E-mail address:* jacristanchoc@unal.edu.co (J. A. Cristancho).

## 1. Introduction

Due to the uneven distribution of lightning activity on the planet, global lightning maps show that some places in the world have very high levels of lightning activity. In these places, the population is more prone to suffer injuries caused by lightning. The top 100 places of the world with the highest lightning activity have a lightning density in a range between 83.45 and 232.52 flashes km-2 yr-1 (Albrecht et al., 2016). Several fatal accidents, reported by both media and technical literature, show that people involved in outdoor activities or placed in open fields are more vulnerable to the risk of lightning (Cristancho et al., 2017; Cooper et al., 2016; Navarrete-Aldana, 2014; OSHA-NOAA, 2016; Walsh et al., 2000).

In order to reduce the risk of death caused by lightning strikes on living beings and damage in structures, buildings, installations and their contents (IEC 62305-1, 2010), several techniques intended to intercept direct lightning flashes and to conduct and disperse their currents safely into the earth have been developed (Bouquegneau, 2010). To prevent the mentioned lightning threats, a lightning protection system (LPS) should be able to capture and divert safely the hazardous impulsive lightning currents into the ground. For fixed external building protection, the IEC62305 and NFPA 780 standards (IEC 62305-1, 2010; National Fire Protection Association, 2017) recommend the use of some materials and components that in most of the cases are heavy and bulky. Nevertheless, in some outdoor, faraway, backcountry or not easily accessible locations, the use of weighty materials and equipment is almost impossible, constraining the possibility to have acceptable LPSs. This condition makes some human activities to be permanently exposed to the catastrophic risk of lightning.

In the research field of technical and industrial fabrics, the electroconductive fabrics (ECF) show great potential of application on several industries (Grancarić et al., 2018; Miao & Xin, 2017; Smith, 2010) but until now, there is a lack of information about their use in power or lightning applications. The electrical signals applied on ECF are low power rated, such as in electromagnetic interference (EMI) shielding applications (Baltušnikaitė et al. 2014) or in strain sensing applications (Wang et al., 2014).

In this paper, it will be shown that some ECFs, subjected to high current impulses in the laboratory, could withstand natural atmospheric lightning impulse currents (LIC) enabling its use in personal lightning protection and power applications. Electrical tests and measurements were performed on some electroconductive textiles, as well as microscopy observations to analyze the behavior of the fabrics against high currents. It is proposed a possible alternative use of some ECFs as components of a lightweight LPS technology for use in tents, portable lightning shelters, and other power applications, stablishing some parameters that limits the current or energy level in the used textiles.

### *1.1. Overview of electroconductive fabrics*

ECF are fabrics with the ability to conduct electrical currents, using conductive materials in their constituent fibers and yarns. According to the weave pattern they can be woven, nonwoven or knitted. Woven fabrics consist of interlacements of yarns in mutually perpendicular directions, interlacing or interweaving warp-and-weft yarns. On the other hand, nonwoven fabrics are textiles whose structure is produced by the bonding or interlocking of fibers with mechanical, chemical, thermal, or solvent methods, or combinations thereof (Miao & Xin, 2017). Knitted fabrics are made up of a single yarn, looped uninterruptedly, producing a braided appearance. The weave structure, pattern and the number of yarns depend on the fabric type. Only woven and nonwoven ECF were used for this research. Figure 1a shows optical micrographs of four ECF utilized in this research, and Figure 1b a warp cross section of a rip-stop ECF (also used in this work), revealing 48 weft yarns of the textile. Each yarn has an average 15 µm diameter with an outer conductive layer of about 1 µm thickness. Commonly in woven textiles, warp yarns are more stressed than the weft ones since the former are stretched to receive the interlaced ones of the latter.

### *1.2. Sheet resistance*

ECF are not homogeneous structures with isotropic electrical current distribution. Electrical modeling of conductive textiles is complicated, since the material must be treated as a combination of resistors in series and parallel connections considering the interlacing filaments (Banaszczyk et al., 2009). An electrical model of the whole ECF could be treated as a very large-scale symmetric network of basic yarn interconnected sections. The entire model can be reduced to multiple single yarns connected in parallel (Zhao et al., 2016) which in the end can be taken as a uniform conductive sheet. The electrical resistance of a full symmetric grid corresponds to that of the basic electric model, which mainly depends on the material, pattern, and geometry of the conductive elements. The electrical resistance value will also depend on several external factors, such as temperature, humidity, pressure or extension (Banaszczyk et al., 2009; Varnaitė-Žuravliova et al., 2016).

Considering the conductive textile surface as a thin film conductor (Zhao et al., 2016), the sheet resistance is frequently used to refer to the electrical resistance of a full symmetric grid, i.e. a square of fabric independent of the side length. This is different from the bulk resistance used for wires conductor specification. Considering the regular three-dimensional resistance $R$ of a conductor, Equation 1 describes the mathematical model of the resistance for a conductor square of sides $W$ and thickness $t$, in the direction of a current $I$, going through the cross-sectional area A.

$$R = \rho \frac{Lenght}{Area} = \rho \frac{W}{t.W} = \frac{\rho}{t}\frac{W}{W} = \frac{\rho}{t} \quad [\Omega] \qquad (1)$$

Eq. 1 is valid only for a square of film conductor, where ρ is the resistivity and $\frac{\rho}{t}$ is the so-called sheet resistance Rsh, then R is independent of the size of the squared sample (Banaszczyk et al., 2010). Despite the unit of electrical resistance in the international system of units (SI) is the ohm (Ω), in the conductive films industry it is commonly expressed in units of ohm per square (Ω/□) which expresses that the resistance is measured on a square of material of side $w$ and it is independent of its side length (ASTM D4496−13, 2013; ASTM F390-11, 2011; Smith, 2010).

### 1.3. Overview of lightning impulse tests

A lightning flash to earth is an electrical discharge of atmospheric origin between cloud and earth consisting of one or more strokes (IEC 62305-1, 2010). Cloud-to-ground lightning effects depend mainly on the high current amplitudes developed when the unbalanced opposite charges between clouds and ground are neutralized. When lightning strikes the ground directly or through a tall object, the currents are radially distributed on the earth in all directions from the striking point to the exterior. Despite the fact that lightning currents could reach more than 200 kA, for negative first strokes the global lightning median return-stroke peak current is about 30 kA and typically 12 kA for subsequent strokes (CIGRE WG C4.407, 2013). For some countries such as Brazil, Colombia and Zimbabwe (former Rhodesia), the average lightning peak value is about 42 kA (CIGRE WG C4.407, 2013; Rojas et al., 2017). Lightning current density diverted into the ground is still hazardous to life and health even at distances of several meters from the strike point, although it is inversely reduced to the square of the distance.

Impulse voltages with front durations varying from less than one up to a few tens of microseconds are considered as lightning impulses (Kuffel et al., 2000). To assess the effects of lightning currents over equipment and materials, standardized lightning-like impulse currents (LICs) are generated in laboratory to simulate natural lightning impulses.

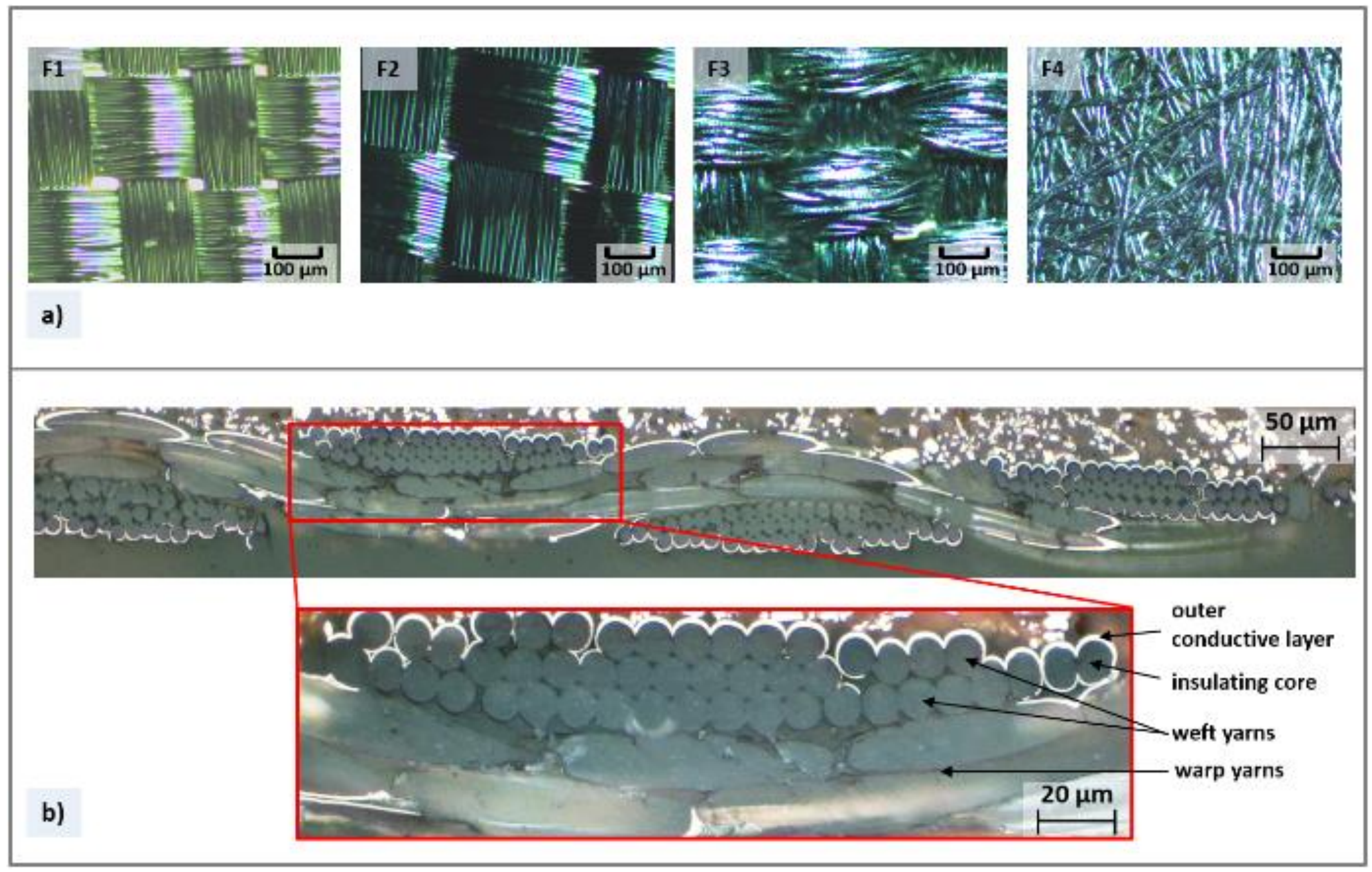

Figure 1. Optical micrographs of electroconductive fabrics were tested in this research.
a) Surface sample of four electroconductive fabrics where the color difference between F1 and other ones is due to an illumination effect; b) warp cross-section of an electroconductive rip-stop fabric showing 48 weft yarns cramped along warp filaments, where in the zoomed image it is possible to identify some inner insulating fibers with their outer conductive shell.

### *1.4. Parameters of lightning impulses currents (LICs)*

The peak value *I*, the front time $T_1$ and the time to half value $T_2$ are basic parameters of the standardized wave shape typically used in laboratories to simulate the effects of LICs on lightning protection systems – LPS – components (IEC 62305-1, 2010) as shown in Figure 2a. Commonly, in a lightning impulse the time $T_2$ is less than 2 ms. For example, a current waveform 4/10 µs of 20 kA (or 20 kA 4/10 µs) refers to a surge of current of 20 kA peak with 4 µs of front time and 10 µs of tail time up to its half value.

When an electrical current flows through a resistive material, heating is produced due to the joule effect. To estimate the energy dissipated by the ECF as joule heating due to each applied current pulse, the instantaneous power applied over the fabric should be considered. The following Equations 2 and 3 are used to evaluate the energy after each current impulse:

$$P(t) = V_e(t) \cdot I(t) = (I(t) \cdot R) \cdot I(t) = I^2(t) \cdot R \qquad (2)$$

$$E = \int_{t=0}^{T} P(t)dt = R \cdot \int_{t=0}^{T} I^2(t)dt \qquad (3)$$

Where *P* is the power, *Ve* is the measured voltage between electrodes, *I* the current applied, *R* the resistance of the conductive fabric, *E* the energy, and *t* is the time.

## 2. Materials and methods

### *2.1. Fabrics used as samples under test*

In order to determine the structural characteristics that provide conduction of high current levels, the objects under test (OUT) were samples of electroconductive textiles with different weave types and manufacturing techniques: rip-stop (named as F1), rip-stop with a side layer coated with a flame-retardant UL94V-0 composite (F2), plain-weave (F3) and a nonwoven textile (F4). In addition to these, a conductive polymeric film (F5) was tested. F1 to F4 are electroconductive fabrics (ECF) produced with yarns of the same characteristics by the same manufacturer. The material F5 is a polymeric film impregnated with carbon black from other industry. Fabric types F1 to F4 were selected because the material type and weave pattern are easily available from different manufacturers in the market, broad commercial offer, low price (comparatively with other ECF), and wide use in electromagnetic shielding. Despite the material F5 is not a textile but a conductive film, it was tested to assess their ability to withstand lightning-like currents and compare it to that of conductive textiles.

The structure of each yarn of the textile ECF follows the structure observed in Figure 1, made from fibers of an insulating polymeric core covered with a conductive material layer. The inner fiber is a monofilament made of polyester, constituting 65% of the entire yarn. The manufacturer data sheet gives the composition of the remaining 35% outer conductive layer as 57 % Cu - 43 % Ni alloy. The UL94V-0 level flame-retardant coating composite of the fabric F2, covers one side of a conductive rip-stop fabric of the F1 type. The main characteristics of the ECF samples are summarized in Table 1. The presented sheet resistance unit is given in ohm per square (Ω/□).

An optical microscope Olympus BX-41 with U-CMAD3 was used to observe the samples before all the tests. Also, a scanning electronic microscope SEM Tescan Vega 3 SB, with secondary electrons (SE) and backscattering electrons (BSE) detectors, was used to observe some selected sections. In addition, the samples were weighed in an analytical Sartorius Entris 224I 1S balance scale.

### *2.2. Lightning impulse current generator setup*

The LICs were generated by lightning impulse current generators (LICG). 8/20 µs current waveforms generated by the LICG were used (T1 of 8 µs and T2 of 20 µs). High amplitude current impulses with that waveshape are considered standardized lightning impulse currents.

The schematic setup of the LICG is represented in Figure 2b. By changing the distance between spheres of the spark-gap, it is possible to control the LICG charging voltage, hence the peak amplitude of the applied current pulses.

Current is measured with a Rogowski coil with integrator, and voltage with a high voltage probe with a 1000:1 ratio. To record these signals, the two channels of an Agilent DSO6104A oscilloscope with ground isolation are used. The ECF samples as objects under test (OUTs) are fastened by two copper clamp electrodes to provide physical support and electrical contacts between the current generator and the fabric sample, as shown in Figure 2b. In this setup, the OUT must have enough low impedance to allow the conduction of the current impulse mainly because the resistance of the fabric sample is part of the LICG discharge circuit.

### *2.3. Procedure*

The procedure for testing the ECFs against lightning currents is depicted graphically in Figure 2c, conducted by means of a LICG over three samples of each four ECFs and one polymeric film similar to the methodology presented in (Cristancho et al., 2018). Three squared pieces of 10 cm x 10 cm of the five materials indicated in Table 1 were trimmed identifying their warp orientation (textiles F1 to F4 and the film F5). Each sample was named and marked according to its material followed by sequential number, e.g. sample F2-3 corresponds to sample 3 of the fabric sample F2, and F5-1 corresponds to sample 1 of polymeric film. Optical microscopy and weighing of the samples were done.

Each sample, placed lengthwise to the surface of the fabric in the warp direction and connected to the circuit through copper clamp electrodes, was subjected to a set of at least four subsequent LICs, starting from 5 kA, increasing in leaps greater than 2 kA with time intervals of 5 minutes between each test. Fabric samples are tested with consecutive 8/20 µs LICs, starting from 5 kA in increasing leaps estimated to be larger than 2 kA, up to the maximum current that the generator can drive with the actual impedance given by the fabric sample.

To observe and compare the impedance behavior of each ECF sample, it is measured and recorded both voltage and current signatures across the OUT on the clamping electrodes.

To investigate the variation of the conducting textile's resistance after the application of successive LICs, this sub-procedure is followed: from the first 5 kA LIC recorded data, we only take the first cycle of both the measured voltage drop and the current. The maximum current value and its corresponding voltage datum are selected from this data set and the dynamic resistance is calculated by means of Ohm's law. The same sub-procedure is followed for each consecutive current test of each fabric sample. Then, it is calculated and tabulated the pseudo electrical resistances, and generated plots of voltage vs. current. In addition, the energy of the first positive semi-cycle of each applied LIC is estimated using Equations (3) and (4) and the energy vs. current characteristic is plotted.

The polymeric film (F5) was also tested in a similar way but only with a first single LIC of 5 kA due to it igniting and melting along the test electrodes contacts before allowing other trials. For this reason, F5 is not studied like the textile samples and it does not appear in most of this paper.

After the LIC test, the samples were observed in the microscope and weighed again to compare the fabrics before and after the tests and to analyze how the fabrics were changed by the high current flow. Additionally, some small sections trimmed out from the samples were examined through the SEM Microscope, with an acceleration voltage of 20.0 kV, to increase the observation capacity for identify the structure of involved materials, and to get more details of the changes over the fabric surface. To estimate the change in the elemental composition of the surface, Energy Dispersive Spectrometry (EDS) analyses were performed over a sample of rip-stop conductive fabric (F1), before and after lightning tests.

Table 1. Parameters of the tested electroconductive fabrics and film.

| Item | Unit | F1 | F2 | F3 | F4 | F5 |
|---|---|---|---|---|---|---|
| Weave pattern | type | Rip-stop | Rip-stop with FR [a] | Plain | Nonwoven | polymeric film |
| Areal density[b] | $g/m^2$ | 90.0 ± 10.0 | 245.0 ± 10.0 | 130.0 ± 10.0 | 90.0 ± 10.0 | 238.0 ± 10.0 |
| Thickness[b] | mm | 0.10 ± 0.01 | 0.16 ± 0.02 | 0.15 ± 0.02 | 0.08 ± 0.01 | 0.10 ± 0.01 |
| Sheet resistance[b] | Ω/□ [c] | ≤ 0.05 | ≤ 0.07 | ≤ 0.07 | ≤ 0.05 | < 31000 |
| Resistivity[b] | µΩ·m | ≤ 5.0 ± 0.5 | ≤ 11.2 ± 1.4 | ≤ 10.5 ± 1.4 | ≤ 4.0 ± 0.5 | $< 3.1\times10^{6} \pm 0.1\times10^{6}$ |
| Conductive material | type | Ni-Cu | Ni-Cu [a] | Ni-Cu | Ni-Cu | impregnated with carbon black |
| Core material | type | polyester | polyester | polyester | polyester | polymeric-film |

(a) Uses a side with a coating layer of flame-retardant level UL94V-0 composite.
(b) The value after ± sign is the specification limit, and after ≤ or < sign indicates the upper limit of the specification.
(c) SI unit is the ohm, but the common unit used for the sheet resistance is the "ohm per square = Ω/□".

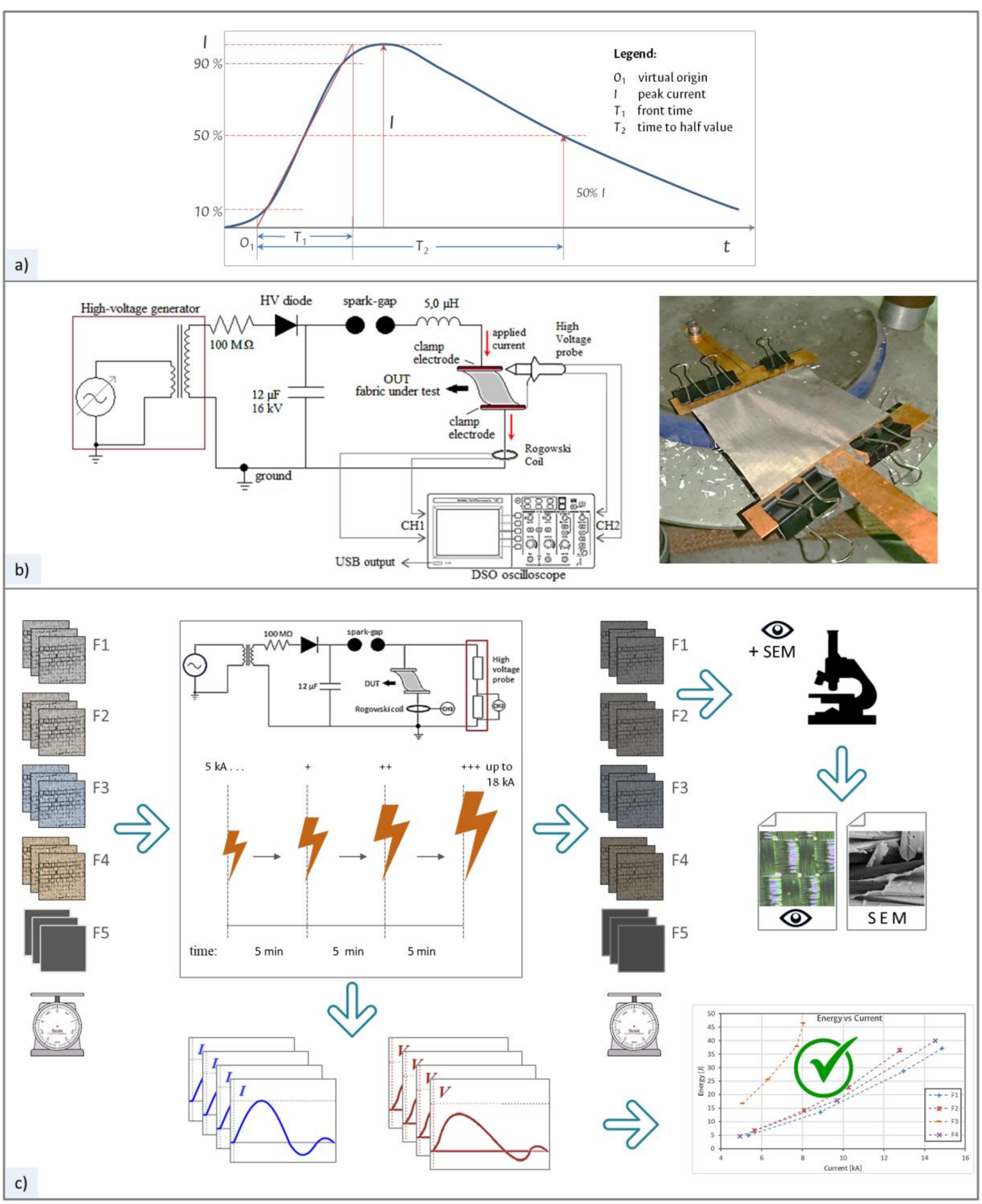

Figure 2. Experimental setup. a) Definitions of a laboratory standard lightning impulse current (LIC) signal with its parameters according to IEC 62305-1; b) test bench setup schematic of the lightning impulse current generator (LICG) used and a sample of electroconductive fabric fastened by the two copper clamp electrodes; and c) procedure for LIC tests over electroconductive fabrics and film.

## 3. Results and discussion

### 3.1. Optical microscopy before tests

Figure 1a shows micrographs of the untested textile samples. F1 and F2 correspond to rip-stop type, F3 to plain-weave type and F4 to nonwoven conductive fabric type. F2 has the back side coated with a flame-retardant composite that increases the temperature strength, as well as the fabric thickness, stiffness, and weight. The picture of the coated polymeric film F5 is not included in Figure 1 since its homogeneous surface image does not provide significant information.

### 3.2. Lightning impulse current (LIC) tests

Figure 3 shows representative samples of each one of the ECF and the coated polymeric film after the application of the last LIC. Regarding the F5 polymeric film (Figure 3q), with the first 5 kA test, it ignited and melted along a test electrode contact, as it can be seen in Figure 3q, avoiding further testing and thus no data results can be provided.

The current and voltage signatures of the first and the last LIC applied to one of the samples of each of the four studied textiles are shown in the upper part of Figure 3, beside each picture. Figs. 3b, 3e, and 3h correspond to the first impulse; Figs. 3 c), 3 f) and 3 i) to the last impulse. Figure 3k is the applied current signal to the F4-3 sample (i.e. the sample 3 of fabric 4) while Figure 3l is the voltage drop across it, shown inverted due to the measurement system connection. After applying the LICs on each test sample, some patterns appeared on the fabric surface, particularly close to the electrodes as shown in Figures 3a, 3d, 3g and 3j. Horizontal patterns perpendicular to the current direction can be noticed in these pictures, the latter indicated by yellow arrows.

For all textile samples of each type, the first cycle peak current value and its corresponding voltage datum of each LIC test were plotted giving a current-voltage (I-V) characteristic as shown in Figure 4 a) to d). These characteristics correspond to the "dynamical" response of the material to the crossing current.

For the first three impulses of all the current-voltage plots presented in Figure 4, an almost linear behavior is noticed, which indicates a constant electric resistance value for those levels of the current. Therefore, the "dynamical resistance" of the ECFs for the first impulses was calculated using the I-V data, as it is shown in the plots of Figure 3 (b, e, h and k), as well as Ohm's law. The average resistance values obtained for the textiles in ascending order are: 15 mΩ for F1, 19 mΩ for F2, 30 mΩ for F4, and 57 mΩ for F3.

The application of the fourth LIC produced a considerable variation in the electrical behavior on all tested ECF, particularly for the F3 plain-weave fabric. The most evident change observed in the sample F3-2 shows an erratic behavior compared with the other three samples. On the other hand, for the F4 fabric, the last applied LIC of 19 kA, produced changes of about 33% and up to 96%. This large variation was not expected from the projections of the previously three measured values of the textile electrical resistance.

### 3.3. Energy estimation

The energy applied by the first positive semi-cycle of each LIC was estimated using Equations (2) and (3). The average resistance obtained from the first three LICs of Table 2 is considered as the fabric resistance. The energy in joules related to the applied current in kiloamperes on each weave pattern is plotted in Figure 4e.

After the second applied LIC, a progressive slight change in the color of the fabrics was noticed. From the original silver color, the surface acquires a soft dark tone after each impulse. This suggests overheating of the material due to the joule heating effect produced by the application of high LICs, which could even sublimate the thin conductive metallic layers when the current density surpasses a critical value. LICs applied with enough high amplitude can produce loss of conductive material layer in some areas changing the fabric structure. This phenomenon increases the ECF electrical resistance and therefore the dissipated energy in the conductive material when a successive impulse is applied. In a previous work, the critical LIatamplitude for the sublimation and break-down of a dog-tag metallic chain (i.e. pearl-like necklace) was estimated at 17.4 kA peak for a 8/20 µs waveform (Latorre et al., 2016). For the tested textile samples, the higher the lightning current impulse amplitude, the more the effect on the fabric surface was observed.

The energy vs current characteristic (Figure 4e) shows an evident difference for the plain-weave F3. This plot reveals a similar behavior for the fabrics F1, F2 and F4, and that the plain-weave electroconductive textile F3 dissipates more energy, even at lower current values. This is consistent with their electrical resistance presented in Table 2, where for F3 it presents the highest value of the four textiles evaluated. Fig 4 also shows a clear particular behavior for F3 (Figure 4c) after the third LIC, indicating a strong change in its electrical resistance and conduction mechanism.

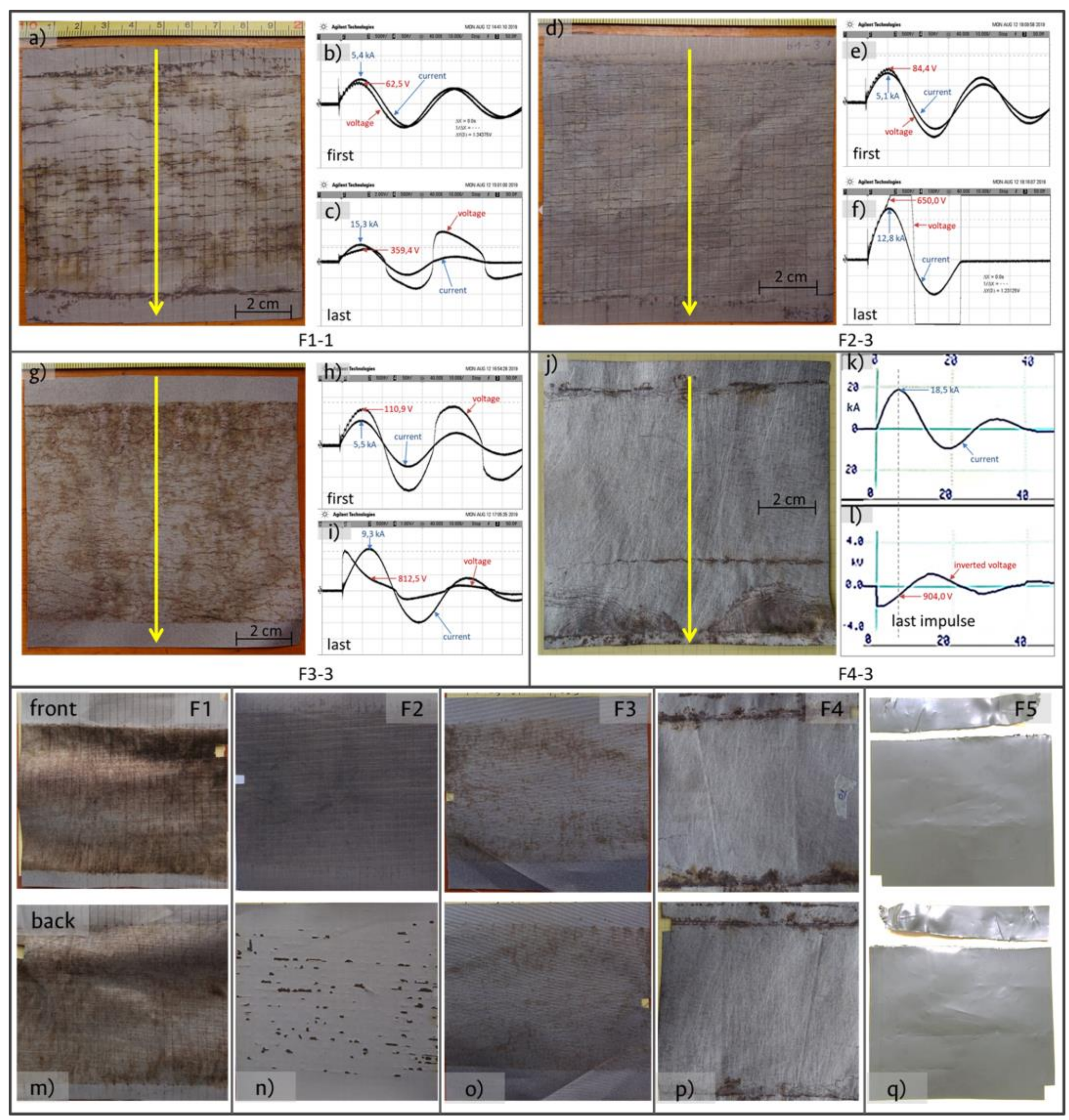

Figure 3. Photographs of tested electroconductive textile samples showing some current and voltage signals. Images of rip-stop F1-1 (a), rip-stop with flame retardant (d), plain-weave (g) and nonwoven (j) conductive fabric samples after lightning impulse current (LIC) indicating the direction of the current flow with yellow arrows. Next to them are both current and voltage signatures of the first and the last impulse as follows: plots (b), (e), and (h) correspond to the first applied impulses, while (c), (f), and (i) to the last ones. For the nonwoven fabric sample F4-3, the current (k) and voltage (l) signatures are shown separately, the latter with negative polarity by the acquisition system connection. After LIC tests, some marks particularly perpendicular to the current flow appear in the fabric sample surfaces. In addition, both sides of each tested electroconductive textile samples F1 (m), F2 (n), F3 (o), F4 (p) and the polymeric film sample F5 (q) are shown after the last applied impulse. The textiles (samples F1 to F4) show marks in the front and back surfaces and the film (sample F5) shows the cut caused by the melting of the electroconductive surface. Note the flame-retardant composite in the back surface of the F2 (n) with a scratch pattern that is perpendicular to the direction of the current (vertical for all pictures).

Table 2. Average resistance in mΩ for the first three lightning current impulses of each fabric sample, average for fabric type, and expanded uncertainty for 95% of confidence for each weave type.

| Sample | F1-1 | F1-2 | F1-3 | F2-1 | F2-2 | F2-3 | F3-1 | F3-2 | F3-3 | F4-1 | F4-2 | F4-3 |
|---|---|---|---|---|---|---|---|---|---|---|---|---|
| Resistance [mΩ] | 15 | 15 | 16 | 15 | 17 | 24 | 56 | 47 | 68 | 29 | 28 | 34 |
| $R_{average}$ [mΩ] | | 15 | | | 19 | | | 57 | | | 30 | |
| Exp. uncert. [mΩ] | | 1 | | | 5 | | | 12 | | | 4 | |

### 3.4. SEM observations

To assess the morphology changes on the OUT fabrics after the LICs, some selected samples were observed by means of the scanning electronic microscope (SEM). The magnification was increased over some scratches on the surface as shown in Figure 5.

The secondary electron image detection (SEI) of the SEM microscopy (Figure 5m and 5n) allows easy identification of the conductive and non-conductive materials, mainly because the atomic number of metal atoms are greater than those of isolating organic compounds. The dark areas of the non-conductive polymer filaments contrast strongly with the clear areas of the conductive metallic coating.

Under the conditions described, SEM micrographs show optically that the woven fabrics (samples F1, F2 and F3) would better support the high currents flow than the nonwoven one (F4). F4 nonwoven fabric surface (Figure 5j, 5k, 5l) has regions that endure worse the high-currents flow despite their electrical response and visual aspect.

This might suggest that the woven fabrics have a better performance to endure the current flow, i.e. the most ordered and straight the weave the more its capacity to drive impulsive currents. However, this observation could depend on the length, geometry, the material and the production technique of the fibers in nonwoven fabrics.

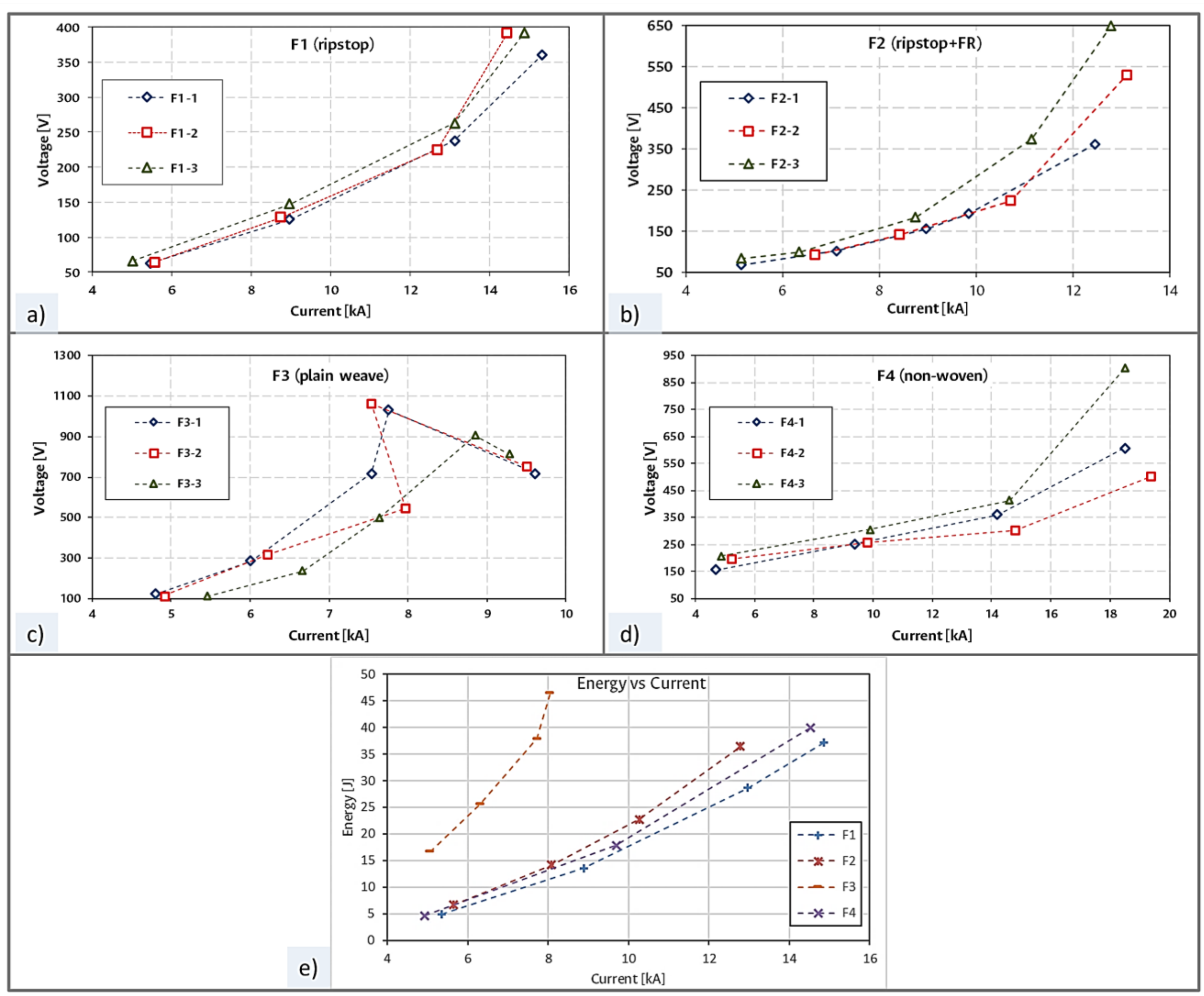

Figure 4. Voltage-current graphs of the LIC tests on the electroconductive fabrics F1 to F4 (a to d) and energy-current averages plot of the subsequent LICs tests over the fabrics (e). Three samples of each fabric were tested with four subsequent increasing LICs for samples F1 (a), F2 (b) and F4 (d), and five impulses for sample F3 (c), where the sequence tests are represented by the colored lines (blue ◊, red □, green Δ), one for each sample. Note that horizontal and vertical scales are different in all graphs.

Furthermore, the F3 plain-weave with a pattern like that of the rip-stop fabrics F1 (Figure 5a, 5b, 5c) and F2 (Figure 5d, 5e, 5f) but with a loose weave, shows a greater resistance and consequently greater energy dissipation. Despite similar research is needed over knitting fabrics and this weave type is not treated in this work, considering they have a more intricated weave pattern that changes the yarn paths, lower performance would be expected.

With the SEM energy dispersive spectrometry analysis (EDS), the estimation of the elemental composition of the conductive surface layer was performed over a sample of F1 rip-stop fabric. The EDS semi-qualitative chemical analysis identified a composition of 55% Cu - 45% Ni over a non-affected spot area of the fabric surface, as shown in Figure 5m. In the same way, over the observation region of the surface of the sample F1 after lightning current tests, it was identified an elemental composition of 44% Cu - 36% Ni - 16% O - 4% C, as shown in Figure 5n, where carbon and oxygen are two basic elements of the inner polyester filaments.

The resistance value of the electroconductive fabric is a key parameter of the ECF intended to drive lightning currents. The elemental composition of the yarns conductive layer defines the resistivity, therefore the resistance. According to the SEM-EDS analysis (Figure 5), the tested fabrics in this research are made from yarns with a conductive layer of copper-nickel alloys in a concentration of approximately 55% Cu and 45% Ni. Bundling many yarns from materials with enough low resistivity can lead to weaves with low resistance able to withstand lightning currents.

Carbon fibers, carbon nanotubes CNT, graphene-coatings, and other non-metallic conductive materials based on organic compounds have relatively high resistivities compared with metallic ones, such a copper, leading to high resistance values and therefore, high energy dissipation in form of heat. The polymeric carbon-coated film F5 shows that it cannot withstand even a single 5 kA LIC, because the heat generated melts the material itself along one of the contact electrodes.

### *3.5. Loss of material and “crosswise asperities”*

Table 3 shows the weights average taken for the three fabric samples before and after tests, revealing an average loss of approximately 0.8 % for F1 (ripstop), 23.2 % for F2 (ripstop + flame retardant), 0.4 % for F3 (plain-weave), and 0.7 % for F4 (nonwoven). Because of the melting and sublimation of the outer metallic layer of the yarns, the conductive material is lost in some regions of the fabric surface. In this case the polyester base material of the filament core is exposed and can be recognized as some dark patterns presented in previous figures. This loss of material results in mass loss.

In particular, the F2 samples show the greatest loss, also due to the release of small pieces of the flame-retardant layer composite, following the pattern of scratches left on the ECF surface as shown in Figure 3n. The final damage of the flame-retardant coating composite on the back side, and the surface marks on F2, suggest that this composite help to increase the physical strength of the fabric against lightning current damage, possibly due to the increased capacity of the coating to support higher temperatures. However, the composite layer could increase the electrical resistance compared with sample F1, as shown in Figure 4 and Table 2, because it decreases the heat release capacity of one side of the fabric. Additionally, the composite layer increases the stiffness, mass and thickness of the ECF, hindering its manageability.

As shown through micrographs of Figs. 1 and 5, the metal layer on the yarn of the fabric forms a net of tubes that can conduct LICs. This conductive network may gradually be lost by melting due to the intense currents. Despite this, the fabric continues conducting subsequent lightning impulse currents through the remaining paths but losing more of the metal coating, in some cases until the non-conductive inner polyester core is exposed as shown in Figure 5. The SEM micrograph of Figure 5l even reveals molten polyester of the core outside the little tubes formed by the conductive sheath of the nonwoven fabric. The mass change, consequence of the melting and loss of material after lightning tests, presented in Table 3, does not suggest any relation with the capacity of the electroconductive fabric to withstand the laboratory lightning currents. The greatest mass loss (23.2 %) was in the F2 rip-stop fabric with flame-retardant layer and the lowest loss (0.4 %) in the F3 plain-weave, the latter with the worst electric result. Thus, this result can be related most with the mechanical effect and would only indicate the loss of material shot out in the air, suggesting that the lost material does not necessarily go out of the weave and could remain tangled up in the fabric yarns, as shown in Figure 5 (c, e, f, i, and l).

The horizontal scratches, perpendicular to the current flow, appear because of the overheating of metallic outer layer of the yarns by the joule effect. Even though it is not totally clear how the horizontal pattern is produced, it is possible to give an explanation by thinking of the electroconductive tested fabrics as networks of conductive tubes. When a lengthwise current flows through a single tube, and this tube experiences a change in the direction, the current density inside increases or decreases according to the change which is subjected to a pipe that transports liquids. Where the current density increases, the heating increases because of the joule effect and, conversely, when the current density decreases, the heating decreases.

As the fabric has a regular symmetrical array, the pattern of scratches appears crosswise to the current flow. This agrees with the observation of the previous paragraph that relates ordered and straight weaves with increased capacity to drive impulsive currents. According to this, it is possible that a perfectly smooth, plain, and straight nonwoven fabric (similar to F4) with most long fibers oriented lengthwise to the current flow could withstand higher electric currents despite their fragility. Therefore, both the weave pattern and the fabric structure have an important effect on their response against LICs.

Considering the fabric as a simple homogeneous conductive surface, the loss of material, due to the conduction of a high lightning current, leaves some scratches or "crosswise asperities" that constrain the conduction of the electrical current to certain remaining paths, in similar way to the interface between two bulk electrical contacts. Figure 6 shows a sequence representation of the crosswise asperities on the conductive textile surface. The remaining "contact paths" after the LIC test can provide the conducting routes for the electrical current flow of subsequent discharges, as lightning flashes could have more than a single stroke (Rakov et al., 2013).

Crosswise asperities produced after tests on the conductive surface of the fabrics maintain the ability to conduct electrical currents but increase the resistance of the material, as shown in Figure 4 for the four fabric types, where its resistance value has a considerable change after the third lightning impulse current. For these fabrics, ohmic behavior can be considered until the third LIC whose current value depends on the fabric itself.

However, the sample F3-2 (sample 2 of the fabric F3 – plain-weave) shows a strong variation after the 8 kA impulse, reducing the current value but increasing the voltage. At 8 kA gives a resistance of 69 mΩ and for the next peak value of 7.5 kA gives a resistance of 141 mΩ. A higher current value was expected at this point, but the output current of the generator setup is determined by the impedance of the fabric. This unexpected point in the F3-2 response shows that the F3 fabric has a critical value of current conduction since it also occurred for F3-1 and F3-3 despite the little shift of their response.

For subsequent discharges, the scratch is in itself an interface between neighbor conductive areas with discontinuities or gaps that determines the paths that can effectively conduct subsequent currents. As it is represented in Figure 6c, the remaining conductive paths continue driving currents, while open ways and gaps cannot do it.

However, in absence of these remaining conductive paths and with enough high electric field, the air dielectric strength of the gap distance between conductive meeting areas breaks down, conducting the currents through the surface. The dielectric strength of air depends on the shape and size of the electrodes, temperature, pressure, and moisture of the air. Under homogeneous conditions, at atmospheric pressure, and in accordance to Paschen's law for air, the breakdown voltage of a 10 µm gap is less than approximately 350 V (i.e. 35 V/µm) (Peschot et al., 2014).

These changes suggest that there is more than one conduction mechanism in the ECF.

Table 3. Mass loss in 10 cm x 10 cm electroconductive fabric samples before and after tests.

| Samples | Before | | After | | Difference | % |
|---|---|---|---|---|---|---|
| | [mg] | Std. Dev. | [mg] | Std. Dev. | [mg] | |
| F1-Avg | 828.3 ± 0.3 | 4.4 | 822.1 ± 0.3 | 6.2 | 6.2 ± 0.6 | 0.8 |
| F2-Avg | 2389.4 ± 0.2 | 52.1 | 1836.1 ± 0.5 | 137.0 | 553.3 ± 0.7 | 23.2 |
| F3-Avg | 1204.6 ± 0.1 | 6.8 | 1199.4 ± 0.6 | 7.0 | 5.2 ± 0.7 | 0.4 |
| F4-Avg | 791.9 ± 0.5 | 29.8 | 786.1 ± 0.2 | 30.1 | 5.8 ± 0.7 | 0.7 |

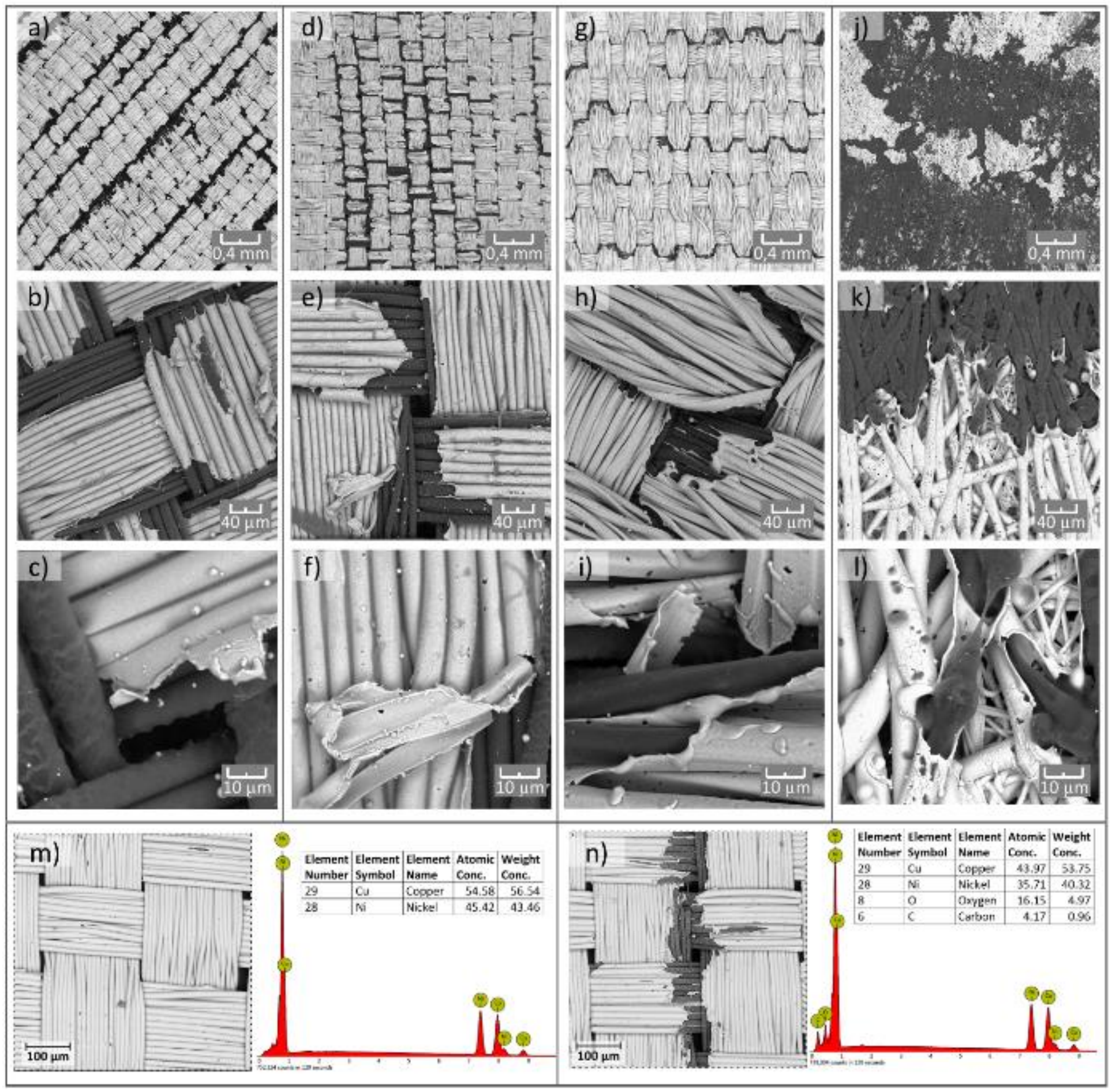

| Element Number | Element Symbol | Element Name | Atomic Conc. | Weight Conc. |
|---|---|---|---|---|
| 29 | Cu | Copper | 54.58 | 56.54 |
| 28 | Ni | Nickel | 45.42 | 43.46 |

| Element Number | Element Symbol | Element Name | Atomic Conc. | Weight Conc. |
|---|---|---|---|---|
| 29 | Cu | Copper | 43.97 | 53.75 |
| 28 | Ni | Nickel | 35.71 | 40.32 |
| 8 | O | Oxygen | 16.15 | 4.97 |
| 6 | C | Carbon | 4.17 | 0.96 |

Figure 5. SEM micrographs and SEM-EDS chemical elemental composition. The SEM micrographs (at HV 20.0 kV) of the four textile samples after the lightning impulse tests show the change in the conductive surface of the four tested types: a), b) and c) rip-stop F1; d), e) and f) rip-stop with a side coated with a flame-retardant F2; g), h), and i) plain-weave F3; j), k), and l) nonwoven type F4; and SEM-EDSs show semi-qualitative chemical elemental composition analyses before (m) and after (n) lightning tests on a sample of the rip stop fabric F1.

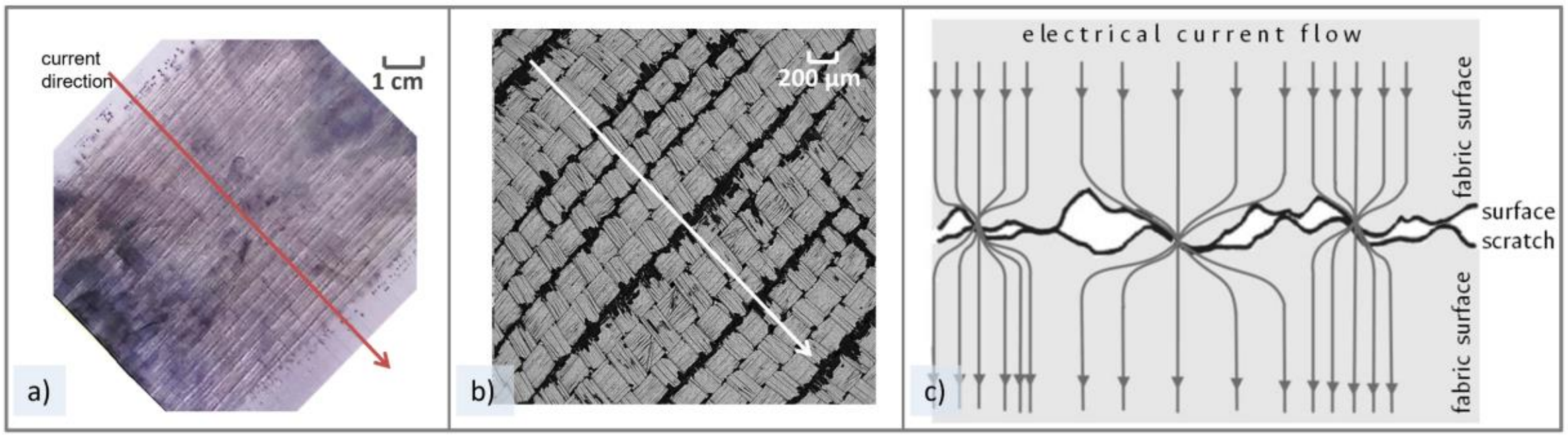

Figure 6. The lightning impulse currents (LICs) produce a scratch pattern on the fabric surface (a) perpendicular to the current direction, easily visible in the SEM image (b). The scratches left on the conductive fabric surface after LIC tests, can be considered as crosswise asperities as shown in the free interpretation of the formed contact interface (c).

### *3.6. Conduction mechanisms and LIC endurance of the tested electroconductive textile.*

The obvious conduction mechanism is the flow of current through the material of the conductive surface itself, as inferred for the first three impulses. For subsequent discharges above the third, the scratches become interfaces between neighbor conductive areas, with gaps and bridges, determining the paths that could effectively conduct new currents. The remaining conductive paths continue to carry new currents. On the other hand, in absence of conductive paths in the crosswise asperities, the electric field can be high enough at the open meeting areas or gaps, breaking down the air strength and producing small discharges (sparks) that bridge the gaps, enabling the conduction of successive lightning currents through the electrical discontinuities.

For fabrics F1, F2 and F4, the resistance behavior displayed in the Current-Voltage characteristics of Figure 4, extends to higher current values suggesting that, at least electrically, these textiles support better LICs than F3. Despite more measurements are needed, the first three impulses tests for all tested fabrics show an approximately linear ohmic resistance behavior (as the voltage follows the current signal and the plots I-V show a somewhat linear trend), after which the yarns change their structure enough to modify permanently the weave, implying an alteration in the electrical current conduction mechanism in a similar way to mentioned previously. From the outcomes of our experimental tests, there is not enough evidence to conclude that subsequent LICs below a critical peak value can produce progressive damage in the fabrics. Conversely, it can be said that after the third current impulse, suggested as a critical value, the scratches are more marked and the damage is progressive, as each time fewer and fewer conductive paths remain.

Woven type textiles comprise weave patterns according to their final use. The samples of the four tested weaves show a tendency to begin to burn up more easily crosswise, perpendicular to the current flow, and left on the surface fabric a pattern of horizontal scratches. In the nonwoven fabric F4, as shown in Figs. 3 and 5, the horizontal melting effect is marked not only as lines but also as large areas following a perpendicular pattern.

The rip-stop fabric has a special weave array with a reinforcing technique that makes them mechanically resistant to ripping and tearing. Conversely, the tested nonwoven fabric appears more fragile with a paper-like brittle feature. Despite the similar energy-current and voltage-current characteristics against lightning impulses between F1 and F4, fragility is an undesirable attribute.

Considering only the visible and electrical effects of the lightning currents on the tested fabrics and discarding the mechanical ones, the rip-stop weave has proven to be a tough material with better performance to support lightning impulse currents like those produced by electrical atmospheric discharges.

Furthermore, considering the lightning parameters given in Tables A.1 and A.3 of IEC62305-1 (IEC 62305-1, 2010), the 50% value of a typical first negative stroke corresponds to a current of 20 kA peak and it has a probability P=0.8. This suggests the possibility to conform lightweight and portable LPS with electroconductive fabrics according to the results of the LIC tests. Additionally, it is very important to remember that both direct and indirect lightning impulse currents from natural atmospheric discharges can cause injuries. In fact, a larger number of fatal accidents are caused by step potentials, despite their relatively low current intensities, but sufficient to produce cardiorespiratory arrest (Cooper et al., 2016). Further study is necessary on this subject to have more accurate information.

## 4. Conclusion

Four types of electroconductive fabrics and a carbon-impregnated polymeric film were tested in high-voltage laboratories to assess the effect of lightning currents on them. Lightning impulse currents (LIC) of 8/20 µs waveshape were generated and used in the laboratory to simulate natural lightning strikes. Consecutive LICs with increasing amplitudes greater than 5 kA, were applied to 10 cm squared samples of rip-stop, plain-weave, nonwoven electroconductive fabrics, and over a polymeric carbon impregnated film. The tested fabric samples withstood the trials with the progressive increasing currents up to 18 kA, but not the carbon impregnated film which could not withstand even the 5 kA current. For each fabric tested, a constant ohmic resistance behavior was observed since the voltage follows the applied current, up to a certain critical value of the applied LIC, after which the current-voltage characteristic changed. Therefore, two mechanisms of electrical conduction on the fabric surface are proposed. The microscopy inspections reveal that the pattern of scratches left by the LICs on the fabric surfaces are due to the melting of the outer conductive layer of the fabric yarns. The outcomes suggested that from the four tested electroconductive fabrics, the rip-stop is the better weave type for potential application in lightweight and portable lightning protection systems to mitigate lightning risks in outdoor applications such as tents and mobile shelters.

## Conflict of interest

The authors do not have any type of conflicts of interest to declare.

## Acknowledgments

The authors would like to thank the Universidad Nacional de Colombia (UNAL) and its professors for all the help and support, access to the SEM and the use of the HV laboratories, Eng. Francisco Amortegui and Eng. Fernando Herrera of the LABE and the Electric and Electronic Department. We also thank Dr. Jorge Ignacio Villa of the Sciences Department of UNAL, to Andres Fernando Gil Plazas from the "Centro de Materiales y Ensayos of SENA Regional Distrito Capital" for the SEM technical support, and to Diana Marcela Forero for her help in the review of the text. We also want to express our gratitude to Zhejiang Saintyear Electronic Technologies Co., Ltd. for the generous samples of conductive fabrics provided for our research.

## Funding

This research is partially supported by the former Administrative Department of Science, Technology and Innovation of Colombia – COLCIENCIAS (currently Ministry of Science and Technology) through the grant of Convocatoria 647 de 2014 and through the Project No.37658/2017 of the *Convocatoria Nacional de proyectos para el fortalecimiento de la investigación, creación e innovación de la Universidad Nacional de Colombia* 2016-2018.